\documentclass[onecolumn]{aastex631}
\usepackage{bbm}
\usepackage{mathrsfs}
\usepackage{amssymb,amsmath}
\usepackage{rotating}
\usepackage{color}

\newcommand\de{\delta}
\newcommand\ep{\epsilon}

\renewcommand\th{\theta}

\newcommand\ta{\tau}

\newcommand\om{\omega}

\newcommand\Om{\Omega}

\newcommand\ie{\emph{i.e.}}

\newcommand\beq{\begin{equation}}
\newcommand\eeq{\end{equation}}
\newcommand\bea{\begin{eqnarray}}
\newcommand\eea{\end{eqnarray}}
\newcommand\bal{\begin{align}}
\newcommand\eal{\end{align}}

\newcommand\fr{\frac}

\renewcommand\bal{\mbox{\boldmath$\alpha$}}

\newcommand\erf{{\rm erf}}
\newcommand\tom{\tilde\om}

\begin{document}

\title{The speed of transmission of phase modulated signals through a plasma medium}

\begin{abstract}
The impossibility of sending pulses of radio waves (Morse codes) through an ionized medium, despite the superluminal phase velocity of the constituent modes, has been demonstrated in many and various ways; essentially the reason is because each pulse, or wave packet, propagate through the plasma at the group velocity, which is subluminal.  Nevertheless, messages can also be encoded as {\it phase} modulations of a monochromatic carrier wave, with more than one constituent modes (which may mathematically be extracted by Fourier transform).  These modes propagate at their respective phase velocities and, upon reassembling them on the receiver's side, can become the original signal with the original message it bore having propagated at the phase velocity of the carrier wave, \ie~superluminally.  We provide a concrete working scenario of transmitting a message for arrival with a time lead (compared to vacuum propagation) which is an order magnitude more than the duration of the message itself.    It is also shown that the distortion of the signal due to the multiplicity of modes induced by the phase modulation is minimal if the bandwidth of the signal, including the duration of its onset and offset, is a small fraction of the carrier wave frequency.  Thus phase and pulse modulations are fundamentally different phenomena in terms of their propagation speeds.
\end{abstract}


\author{Richard Lieu}
\affiliation{Department of Physics and Astronomy, University of Alabama, Huntsville, AL 35899}

\section{Introduction}

The question of the speed of signal communication by electromagnetic wave propagation through a dispersive, absorptive, or gain medium has been a subject of much investigation in the literature, \cite{som14,bri14,war01a,war01b}.  The consensus is that any sharp front edge marking the onset of a pulse propagates through a medium at exactly the vacuum speed of light $c$.  If, on the other hand, there is no sharp edge, the upper limit (or `luminal cap') on the signal speed could be more relaxed, but then one also loses the notion of an exact arrival time of the pulse for comparison with vacuum propagation.  The problem arises because propagation in a medium invariably distorts the pulse profile, even though this does not happen in vacuum, so that a common point on the profile is unavailable for comparison of speeds.  For a symmetric pulse with a well defined peak that manages to maintain its symmetry during passage, the media in question are ones that would not support any superluminal motion of the peak (this is equivalent to saying that the group velocity of the pulse does not exceed $c$, see below).  Thus, one way or another, there is no evidence to date that information communicated in the form of pulses can be transported at superluminal speed.  

For the specific case of propagation in a plasma medium, which is the subject of this paper, the best known example is that of a Gaussian pulse with an underlying carrier wave frequency $\om_0$ exceeding the plasma frequency $\om_p$.  Under the scenario, it is well established that the pulse peak moves at group velocity of the carrier wave, namely \beq v_g = \fr{d\om}{dk} \Big|_{\om_0} = c \sqrt{1-\fr{\om_p^2}{\om_0^2}} < c, \label{vg} \eeq 
and the pulse is broadened while maintaining its Gaussian shape.  Although the phase velocity \beq v_p = \fr{\om}{k} = \fr{c}{\sqrt{1-\fr{\om_p^2}{\om_0^2}}}, \label{vp} \eeq exceeds the vacuum speed of light $c$, this velocity applies to a Fourier mode of frequency $\om$, which is one of many infinite wave trains of equal (and arbitrary) phases comprising the pulse; thus no information can be extracted from any of them.  Instead, any useful information is borne by the pulse, which has a clear time signature indicated by its peak, but which is transported at $v_g < c$.  

In the above discussion one query was not addressed: is it possible to transmit information without enlisting pulses, which are ensembles of Fourier modes of equal phase?  It should be emphasized that such ensembles are not the only way of breaking the symmetry of time homogeneity.  Another way is to assemble modes of {\it systematically varying} phases to embed information in the form of a phase modulated signal.  Such signals have constant intensities, \ie~there are no pulses, yet a phase sensitive detector is able to extract the information it bears.  The potential merit of this approach is clear, while pulse envelopes propagate at the group velocity $v_g < c$ in a plasma, phase variations propagate at phase velocity $v_p > c$ and could in principle be communicated superluminally.  

\section{A phase modulated steady source; Fourier components}

As case in point, the following recipe of an informative message being encoded phase modulated signal is presented.  Consider the voltage (or amplitude) time series comprising a plane monochromatic wave of frequency $\om_0$ propagating along the $z$ direction from time $t\to -\infty$ to $t\to\infty$, except during the interval $0 < t < T$ where $T = 2 N\pi/\om_0$ for some large and positive integer $N$, when the phase of the light waves evolve with time as the sum of $\om_0 t$ and another gently undulating sine wave in $t$.  More precisely, the phase variation is of the form \beq \phi (t) = \om_0 t ~{\rm for}~t \lesssim 0~{\rm or}~t \gtrsim T;~{\rm and}~ \om_0 t + \ep\sin\Omega t ~{\rm for}~0 \lesssim t \lesssim T~{\rm where}~T=\fr{2N\pi}{\om_0}, \label{phi} \eeq where $N$ is a large and positive integer and \beq\ep\ll 1,~\Om\ll\om_0. \label{ep} \eeq  In this way, the voltage time series is expressible as \beq V(t) = \left[{\rm erf} \left(\fr{t}{\sigma}\right) - \erf\left(\fr{t-T}{\sigma}\right)\right] e^{i(\om_0 t + \ep\sin\Omega t)} + \left[1-\erf\left(\fr{t}{\sigma}\right) +\erf\left(\fr{t-T}{\sigma}\right)\right] e^{i\om_o t}, \label{volt} \eeq where $\erf (x)$ is the error function. This gives the voltage at $z=0$, the point of emission of the steady source, for all times $t$.  Note in particular that the source intensity is $I (t) = V(t) V^* (t) = 1$, and is constant at all times $t$, \ie~there are no pulses.

To investigate the effect of propagation through a plasma medium, one must first Fourier transform the voltage time series into modes.  Defining the transform as \beq \tilde V(\om) = \int_{-\infty}^\infty V(t) e^{-i\om t} dt, \label{FT} \eeq one obtains \beq  
\tilde V(\om) = \de (\om-\om_0)+\sum_n 2i J_n (\ep) \left\{ \fr{e^{-(\om - \om_0 -n\Om)^2 \sigma^2/4}}{\om - \om_0 -n\Om}  [-1 + e^{-i(\om-\om_0-n\Om)T}]\right\}  + \fr{e^{-(\om - \om_0)^2 \sigma^2/4}}{\om - \om_0} [1-e^{-i(\om-\om_0)T}] 
\label{FTofV} \eeq where use was made of the relations \beq \int_{-\infty}^{\infty} \erf (t) e^{-i\om t} dt = -\fr{2ie^{-\om^2 \sigma^2/4}}{\om} \label{FToferf} \eeq and \beq e^{i\ep\sin\theta}= \sum_{n=-\infty}^{\infty} J_n(\ep) e^{in\th}. \label{Jake} \eeq
It can readily be verified that (\ref{FTofV}) inverts back to (\ref{volt}) via the standard integral \beq V(t) = \fr{1}{2\pi} \int_{-\infty}^\infty \tilde V(\om) e^{i\om t} d\om. \label{Finv} \eeq If the signal is detected at a distance $z$ downstream and the medium in between is vacuum, the exponent in the integrand will have to be $i\om (t-z/v_p)$ instead of $i\om t$, where $z/v_p$ is the time for each mode to cover the distance $z$ when it propagated at the vacuum phase velocity, (\ref{vp}).

\section{Propagation in a uniform plasma medium}

As noted in section 1, for propagation in a plasma each mode will cover the distance $z$ in time $z/v_p$, where $v_p$ is the phase velocity as given by (\ref{vp}).  The signal at the receiver's end is therefore given by \beq V(t) = \fr{1}{2\pi} \int_{-\infty}^\infty \tilde V(\om) e^{i\om [t-z/v_p(\om)]} d\om. \label{Finv} \eeq To see how this is done, we calculate in some detail the inversion integral (\ref{Finv}) as applied to the second term on the right side of (\ref{FTofV}), since the rest of the terms are dealt with in a similar fashion (inverting the first term is elementary).  To begin with, note that the time of flight $z/v_p$ may be expanded in terms of first order of the small quantity \beq \fr{\tilde\om}{\om_0} \ll 1 \label{tom} \eeq where $\tilde\om=\om-\om_0-\Om$, as \beq \fr{z}{v_p} =t_0 -\ta_0 +\ta (\tilde\om), \label{zvp} \eeq where \beq t_0 =\fr{z}{c};~\ta_0 = \fr{z}{c}\fr{\om_p^2}{2\om_0^2} \ll t_0;~\ta (\tilde\om) = \fr{2(\tom +n\Om)}{\om_0} \ta_0 \ll \ta_0 \label{taus} \eeq On the right side $t_0$ is the vaccum propagation time of all modes, $\ta_0$ is the lowest order correction to $t_0$ due to the phase velocity of the mode with frequency $\om_0$ exceeding the vacuum speed of light $c$, and $\ta (\tilde\om)$ is the next order correction due to the presence of other modes comprising the (information encoded) phase modulation of the monochromatic wave at frequency $\om_0$; these modes perturb $\om$ away from $\om_0$ on either side, by amounts $\tilde\om$ and $n\Om$, with the former related to the width of the error function in (\ref{volt}) and the latter the $\ep$ term there.  The last inequality in (\ref{taus}) is satisfied if (\ref{tom}) holds and $\Om\ll \om_0$ (see below for more about this criterion); while the other inequality holds if $\om_0\gg\om_p$. 

Proceeding, with the help of (\ref{taus}), to invert the second term of (\ref{FTofV}), one obtains \beq \fr{1}{2\pi}\int_{-\infty}^{\infty} -\fr{2i e^{-(\om - \om_0 -n\Om)^2 \sigma^2/4}}{\om - \om_0 -n\Om} e^{i(\om t -\om z/v_p)} d\om = \fr{1}{2\pi} \int_{-\infty}^{\infty} -\fr{2i e^{-\tom^2 \sigma^2/4}}{\tom} e^{i(\tom +\om_0+n\Om) [t-t_0 +\ta_0 -\ta (\tom)]} d\tom \label{intermediate} \eeq  Provided the last inequality of (\ref{taus}) holds, which by glancing at the Gaussian factor in the integrand is the same as requiring that $\sigma\omega_0 \gg 1$, 
the $\ta(\tom)$ term in the exponent of the integrand may be dropped, and the inversion becomes elementary because $t_0$ and $\ta_0$ are not functions of $\tom$.  The result is then, by (\ref{FTofV}) and (\ref{FToferf}), the first term of (\ref{volt}) with the retarded time, or  \beq {\rm time~of~flight}~= t_0 -\ta_0, \label{ToF} \eeq  having been subtracted from the observation time $t$, namely ${\rm exp} [i\om_0 (t-t_0 +\ta_0) +i\ep\sin\Om (t-t_0+\ta_0)] \erf(t-t_0 + \ta_0)$.   The rest of the inversion then proceeds in a similar fashion, resulting in a recovery of the original message, in the form of the phase modulated signal (\ref{volt}), with $t-t_0+\ta_0$ replacing $t$ and with no notable distortion of the signal provided (\ref{ep}) and the last of (\ref{taus}) hold.  
The significance of this result is that the entire message arrives {\it earlier} than vacuum propagation because the propagation time in a plasma is shorter than vacuum by the amount $\ta_0$, see (\ref{ToF}).  This is because, unlike communication by the transmission of pulses which is limited by the group velocity $v_g < c$, phase modulation messaging is limited by $v_p > c$ and can be superluminal.

\section{Parameters for a working model}

Based upon the analysis of the last two sections, one could envisage the following possible scenario of superluminal transmission: plasma frequency $\om_p = 30$~MHz 
(electron density $n_e = 10^8$~cm$^{-3}$), plasma column $z= 3\times 10^{13}$~cm ($\approx 2$ AU), carrier wave frequency $\om_0 \approx 1$~GHz, phase modulation frequency $\Om = 1$~MHz, time for signal onset and offset $\sigma =10^{-6}$~s, and signal strength $\ep =0.1$, and signal duration $T =0.005$~s.  The detection of such a signal is via the imaginary part, because \beq {\rm Re}~(e^{i\ep\sin\Om t}) \approx 1~{\rm and~Im}(e^{i\ep\sin\Om t}) \approx i\ep\sin\Om t. \label{signal} \eeq  This means only $n=\pm 1$ terms in (\ref{Jake}) are needed, and the last inequality of (\ref{taus}) is satisfied in the sense that ratio $2(\tilde\om + n\Om/\om_0)/\om_0$ with 
$n=1$ and $\tilde\om\approx 1/\sigma$ is $\approx 4\times 10^{-3}$, or 0.4 \% which is $ \ll 1$.  This the error in the time of flight advantage $\ta_0$ (compared to vaccum propagation) of the signal, which is given by (\ref{taus}) and has the value $\tau_0 = 0.5$~s for the current scenario.   In other words, the signal arrives earlier than vacuum propagation by a lead time $\ta_0$ which is $\ta_0/T \approx 100$ times than its length $T$.

Turning to the question of detection sensitivity and resolution, the latter is evidently feasible because it is $1/T =200$~Hz which is sufficient to distinguish a line at $\om=\om_0+\Om$ from another at $\om=\om_0$.  On the former, assuming the receiver has a bandwidth $\de\om =10$~MHz, hence coherence length $\approx 10^{-7}$~s, each cycle of oscillation at  frequency $\Om$ would correspond to $N=10$ coherence segments, hence the detection of the phase modulation signal with significance $\ep\sqrt{N} \approx 0.3 \sigma$.  Since the total signal duration $T$ comprise $T\Om\approx 5\times 10^3$ cycles of oscillations at frequency $\Om$, the final detection significance is $\ep\sqrt{NT\Om} \approx 22\sigma$.

\end{document}